\documentclass[sigconf, authorversion=true, nonacm=true]{acmart}

\AtBeginDocument{%
  \providecommand\BibTeX{{%
    \normalfont B\kern-0.5em{\scshape i\kern-0.25em b}\kern-0.8em\TeX}}}

\copyrightyear{2024}
\acmYear{2024}
\setcopyright{acmlicensed}
\acmConference[MSR '24]{21st International
Conference on Mining Software Repositories}{April 15--16, 2024}{Lisbon,
Portugal}
\acmBooktitle{21st International Conference on Mining Software Repositories
(MSR '24), April 15--16, 2024, Lisbon, Portugal}
\acmDOI{10.1145/3643991.3644906}
\acmISBN{979-8-4007-0587-8/24/04}

\usepackage{xspace}
\usepackage{amsmath}
\usepackage{tcolorbox}
\usepackage{tikz}
\usepackage{multirow}
\usepackage{threeparttable}
\newcommand{\urltt}[1]{\texttt{\url{#1}}}
\usepackage{pifont}
\newcommand{\xmark}{\ding{55}}%

\def\girtdata{{GIRT-Data}\xspace}
\def\modelname{{GIRT-Model}\xspace}
\def\dataname{{GIRT-Instruct}\xspace}

\def\numrepoIRT{{50,032}\xspace}
\def\numIRTmarkdown{{98,550}\xspace}
\def\numIRTmarkdownnan{{96,286}\xspace}
\def\numIRTmarkdowndedup{{54,523}\xspace}
\def\numIRTmarkdownlatin{{51,363}\xspace}
\def\numpairsinstruction{{205,452}\xspace}

\def\github{{GitHub}\xspace}
\def\gitlab{{GitLab}\xspace}

\def\Flan{{Flan-T5}\xspace}
\def\participantnum{10\xspace}
\def\humantestnum{40\xspace}

\def\figref#1{Figure~\ref{fig:#1}}
\def\figlabel#1{\label{fig:#1}\label{p:#1}}

\def\tabref#1{Table~\ref{tab:#1}}

\def\tablabel#1{\label{tab:#1}\label{p:#1}}
\def\eqref#1{Eq.~\ref{eqn:#1}}

\def\seclabel#1{\label{sec:#1}\label{p:#1}}
\def\secref#1{Section~\ref{sec:#1}}
\newcommand{\eg}{e.g.,\xspace}
\newcommand{\ie}{i.e.,\xspace}

\renewcommand{\paragraph}[1]{#1}

\newcommand{\bff}[1]{\textbf{#1}}
\newcommand{\und}[1]{\underline{#1}}

\newcommand\circledBody[2][]{{\large \textcircled{\small{#2}}}: #1}

\begin{document}

\title[\modelname]{\modelname: Automated Generation of Issue Report Templates}

\author{Nafiseh Nikeghbal$^\dagger$}\thanks{$^\dagger$This work has been partially carried out as part of the first author's MSc thesis at Sharif University of Technology.}
\orcid{0009-0007-4622-3460}
\affiliation{
  \institution{Independent Researcher}
  \city{Munich}
  \country{Germany}
}
\email{lastname.firstname@gmail.com}

\author{Amir Hossein Kargaran}
\orcid{0000-0001-6253-1315}
\affiliation{%
   \institution{Ludwig Maximilian University}
   \city{Munich}
   \country{Germany}
}
\email{amir@cis.lmu.de}

\author{Abbas Heydarnoori}
\orcid{0000-0001-9785-2880}
\affiliation{%
  \institution{Bowling Green State University}
  \city{Bowling Green}
  \country{USA}
}
\email{aheydar@bgsu.edu}

\renewcommand{\shortauthors}{Nikeghbal et al.}

\begin{abstract}

Platforms such as \github and \gitlab introduce \emph{Issue Report Templates} (IRTs) to enable more effective issue management and better alignment with developer expectations. However, these templates are not widely adopted in most repositories, and there is currently no tool available to aid developers in generating them. In this work, we introduce \modelname, an assistant language model that automatically generates IRTs based on the developer's instructions regarding the structure and necessary fields. We create \dataname, a dataset comprising pairs of instructions and IRTs, with the IRTs sourced from \github repositories. We use \dataname to instruction-tune a T5-base model to create the \modelname.

In our experiments, \modelname outperforms general language models (T5 and \Flan with different parameter sizes) in IRT generation by achieving significantly higher scores in ROUGE, BLEU, METEOR, and human evaluation. Additionally, we analyze the effectiveness of \modelname in a user study in which participants wrote short IRTs with \modelname. Our results show that the participants find \modelname useful in the automated generation of templates. We hope that through the use of \modelname, we can encourage more developers to adopt IRTs in their repositories. We publicly release our code, dataset, and model at \url{https://github.com/ISE-Research/girt-model}.
\end{abstract}





\maketitle

\section{Introduction}

In many popular software repositories, it is common to receive various issue reports on a daily basis. However, when these issue reports lack organization and essential information, it creates complications for issue management and adds to the workload of developers~\cite{bettenburg2008makes, soltani2020significance}. To address this challenge, hosting platforms for software development such as \github and \gitlab have introduced template features for issue reports, commonly referred to as \emph{issue templates}~\cite{GitHub-template-intro} or \emph{issue report templates (IRT)}~\cite{nikeghbal2023girt}. This template feature allows developers to customize the structure of issue reports, specifying the information that contributors should include when opening new issues. Developers on \github can customize IRTs using Markdown or YAML formats, while on \gitlab, customization is done using Markdown. Issue reports written using templates are resolved more quickly and have higher comment coverage~\citep{li2023first}. Additionally, the monthly submission of issue reports decreases after adopting IRTs. However, \citet{nikeghbal2023girt} observed that after analyzing over one million repositories on \github (with more than 10 stars), only nearly 5\% of \github repositories adopted IRTs. This is despite empirical studies showing the benefits of using IRTs, and IRTs were introduced on \github in 2016. A potential explanation for this limited adoption could be developers' unclear understanding regarding which template to employ and what specific information should be included within it. 

\begin{figure}
\centering
\begin{minipage}[htp]{0.47\textwidth}
\small

\begin{tcolorbox}[colback=white, boxrule=0.5pt]
    \begin{tcolorbox}[colback=blue!10, boxrule=0.5pt]
\begin{verbatim}
name: Bug report
about: <|MASK|>
title: <|MASK|>
labels: <|MASK|>
assignees: <|EMPTY|>
headlines_type: # Heading
headlines: ['Describe the bug', 'To Reproduce',
'Expected behavior', 'Screenshots
(if appropriate)', 'Environment', 'Additional 
context']
summary: This issue template is designed to
help users report bugs they encounter while
using our software. Environment section should
asks for "operating system" such as  Ubuntu.
\end{verbatim}

    \end{tcolorbox}
    \begin{tcolorbox}[colback=red!10, boxrule=0.5pt]
\begin{verbatim}
---
name: Bug report
about: Create a report to help us improve
title: '[Bug]'
labels: bug
assignees: ''
---

## Describe the bug
A clear and concise description of what the bug
is.

## To Reproduce
Steps to reproduce the behavior:
1. Go to '...'
2. Click on '....'
3. Scroll down to '....'
4. See error

## Expected behavior
A clear and concise description of what you 
expected to happen.

## Screenshots (if appropriate)
If applicable, add screenshots to help explain 
your problem.

## Environment
- OS: [e.g. Ubuntu]
- Version [e.g. 22]

## Additional context
Add any other context about the problem here.
\end{verbatim}

    \end{tcolorbox}
\end{tcolorbox}

\end{minipage}
\caption[Example of IRT generation for the bug report.]{Example of IRT generation for a bug report. The upper part is the instruction section. If the user explicitly wants a field to be empty, they can add the \texttt{<|EMPTY|>} token. In this example, it is used for assignees. Otherwise, they can add \texttt{<|MASK|>} to let the model decide what to fill in, for example, in the case of about, title, and labels. It's not necessary to fill in any fields, including the summary, but doing so provides more information about what the expected IRT would be.}\figlabel{teaser-example}
\end{figure}

To address this, we propose \emph{\modelname}, an assistant language model tool.
\modelname is capable of generating customized IRTs based on developers' input instructions.
The \modelname could potentially lead to an increase in the adoption of IRTs by assisting developers in customizing their IRTs more easily and faster.

\modelname is designed to function in both modes: when a complete instruction for the model is provided and the user knows exactly what they want, or when only part of the instruction is available and the user is not knowledgeable about IRTs.
The training data for \modelname is derived from existing IRTs employed by various repositories on \github. Our approach to building \modelname is rooted in the concept of instruction-tuning~\cite{mishra-etal-2022-cross, wei2021finetuned, sanh2021multitask}, which involves fine-tuning large language models using instructional data. In our case, the instructional data contains instructions for generating IRTs and the associated IRT pairs. We create a set of instructional data called \dataname. \dataname is constructed based on \girtdata~\cite{nikeghbal2023girt}, a dataset of IRTs. We use both \girtdata metadata and the Zephyr-7B-Beta language model~\citep{tunstall2023zephyr} to generate the instructions. We use the T5-base model with a parameter size of 0.2B as the basis for \modelname. \modelname-base is instruction-tuned using \dataname.

In \figref{teaser-example}, \modelname is illustrated in an example scenario. The instruction part uses the \texttt{<|MASK|>} token when the user does not know about the field, or wants the model to decide, and if the user wants the field to be explicitly empty, they can use the \texttt{<|EMPTY|>} token. We provide a user interface for users to better interact with \modelname (\figref{user-interface}). 

Our main contributions include:
\begin{enumerate}
    \item  We introduce \modelname, to the best of authors' knowledge, the first issue report template (IRT) generation tool. 
    This tool enables users to design IRTs for different issue categories. We also develop a user interface that helps users discover more \modelname capabilities.

    \item We release \dataname, a set of instructional data for the task of IRT generation.
    
    \item In our experiments, \modelname outperforms several baselines (T5 and \Flan) by far in ROUGE (more than 34\%), BLEU (more than 35\%), METEOR (more than 33\%) and human evaluation (more than 48\%).
    
    \item We conduct a user study with \participantnum software engineers and ask them to write short IRTs with \modelname. Participants generally find \modelname beneficial and report great ``usefulness'' and ``goal achievement'' scores. Participants spend less time and generate longer IRTs when using \modelname. The output produced by \modelname does not need to be flawless to be valuable. Software engineers may not always accept \modelname output verbatim but still find it useful in their design process as shown in our user study.
    \end{enumerate}

The rest of this paper is organized as follows.
\secref{background_relatedwork} provides the background and discusses related work.  
\secref{data} introduces our instructional data. This data is used to train \modelname.  
\secref{model} introduces the \modelname, detailing its architecture and fine-tuning.
\secref{evaluations_setup} outlines the experiments designed to evaluate the \modelname.
\secref{evaluations_results} provides the results of evaluations.
\secref{sec-user-study} presents a user study conducted to evaluate the effectiveness of \modelname in practice.
\secref{threats} discusses potential threats to the validity of the experiments and results. \secref{limitation} discusses some of the limitations of the proposed approach.
\secref{conclusion} concludes the paper, summarizes the findings and outlines directions for future research.

\section{Background and Related work} \seclabel{background_relatedwork}

This work is situated within three specific areas: (1) issue report management; (2) issue report/pull request templates; and (3) instruction\allowbreak-tuning of large language models (LLMs). We provide a brief overview of each as a precursor
and motivation to \modelname.

\subsection{Issue Report Management}

Several studies have been conducted to facilitate the more efficient management of issue reports. Since the issue report description is in the text format, most research in this area is conducted in conjunction with advances in natural language processing (NLP). The research in this area includes, but is not limited to, the following: the classification of issue reports \citep{kallis2021predicting, kallis2022nlbse, bharadwaj2022github, wang2022personalizing, izadi2022catiss, colavito2023few, siddiq2022bert}, issue report summary generation~\citep{rastkar2014automatic, ferreira2016bug, mani2012ausum, gupta2021approach}, and issue report title generation~\citep{chen2020stay, zhang2022itiger}. We now discuss each of these research categories in more detail.

\paragraph{\textbf{Classification.}}  The goal of this category of research is to identify the type of each submitted issue report. This would facilitate the management and prioritization of issues. The majority of research in this area categorizes issue reports as a ``bug report'', an ``enhancement/feature request'', or a ``question/support/documentation''
\cite{izadi2022predicting, kallis2022nlbse, kallis2021predicting}.

Most of these studies overlook IRTs and their impact on their research. However, by adopting IRTs, developers can prompt contributors to submit issue reports under the appropriate class type, enabling better and more personalized categorization. This addresses the need to classify issue reports in advance. Furthermore, since IRTs request contributors for more information, 
it could potentially enhance research in this field, resulting in improved classification outcomes.

\paragraph{\textbf{Summarization.}}
Summarizing issue reports is one method to reduce the time and effort required by developers for issue management. \citet{gupta2021approach} developed a two-level method for generating bug report summaries. This approach involves using the title, description, and comments from a resolved bug report to create the summary.

In summarization studies, IRTs are also ignored. According to \citet{li2022follow}'s study, 56.29\% of the IRTs involved developers initially asking users for a summary of the issue. This addresses the need to summarize issue reports in advance.

\paragraph{\textbf{Title Generation.}} The quality of issue report titles is essential for a clear understanding and effective management of GitHub issue reports. \citet{chen2020stay} emphasized the importance of having high-quality issue titles and descriptions. They introduced a technique for automatically creating titles from issue descriptions. 

In another study, \citet{zhang2022itiger} designed a web tool that fine-tuned the BART~\citep{lewis2019bart} model for automated issue report title generation. \citet{zhang2022automatic} apply a similar approach, this time for generating pull request titles. They formulate the task as an automated one-sentence summarization.

In studies on generating titles, researchers often overlook IRTs. However, when using IRTs, developers have the option to set a pre-determined title for the templates. For instance, they might use ``[Bug]:'' as a predefined title.

\subsection{Issue Reports/Pull Requests Templates}

We describe the related work on issue reports/pull requests templates in three aspects: empirical studies conducted, efforts to structure un-templated bug reports, and datasets.

\paragraph{\textbf{Empirical Studies.}}
GitHub introduced pull request templates (PRTs) in 2016, similar to IRTs, with the aim of enhancing the quality of pull requests. \citet{zhang2022consistent} empirically investigated the use of PRTs in 538,864 open-source projects. Their findings indicated that the use of PRTs has a beneficial effect on the maintainability of open-source projects. They discovered that 1.2\% of the repositories use these templates. 

In another study, \citet{li2022follow} performed an empirical investigation of the IRTs/PRTs feature on GitHub. They conducted an empirical study on 802 of the most widely used projects with the goal of uncovering information about the content, impact, and perception of templates. They discovered that these templates typically include a diverse range of components for ``greeting contributors'', ``explaining project guidelines'', and ``collecting relevant information''.  They observed that following the adoption of the template, there is a decrease in the monthly submission of issue reports and pull requests.

Also, \citet{sulun2024empirical} performed an empirical investigation focusing only on IRTs in 100 popular repositories. They observed a growing preference for YAML-based templates in these repositories. They also found projects with a template 
shows significantly reduced resolution time (from 381.02 days to 103.18 days) and fewer issue comments (from 4.95 to 4.32) compared to those without.

\paragraph{\textbf{Structuring Bug Reports.}}  
\citet{song2020bee} introduced a tool named ``BEE'' for structuring and analyzing bug reports. This tool is designed to extract specific elements from bug reports and has the capability to determine the issue type, whether it is a bug, enhancement, or a question. If the issue is identified as a bug, the tool can analyze the natural language text to detect ``observed behavior'', ``expected behavior'', and ``steps to reproduce'' parts of the bug report. These parts are common elements in bug report templates used by the \github community.

In another study, \citet{li2023first} conducted an empirical study on GitHub's bug report templates, exploring their popularity, benefits, and content. They found that bug reports written using templates are resolved more quickly and have higher comment coverage. Additionally, they propose an automated method for converting non-templated bug reports into templated ones, similar to the objectives outlined in \citet{song2020bee}'s study.

\paragraph{\textbf{Datasets.}}
\citet{nikeghbal2023girt} introduced \girtdata, the first and largest dataset of IRTs in both YAML and Markdown formats. This dataset and its corresponding open-source crawler tool are intended to support research on IRT studies. They raised multiple potential research questions when studying IRTs, including how to generate an IRT based on the developers' requirements.

\subsection{Instruction-Tuning}

We consider the related work on instruction-tuning in two aspects: instruction-tuned LLMs and how to generate instructional data.

\paragraph{\textbf{Instruction-Tuned LLMs}.}
LLMs are capable of following general language instructions when fine-tuned with annotated datasets that contain instructional data. Instruction-tuning facilitates the control of LLMs using natural language commands and improves these models performance and generalizability to unseen tasks~\cite{wang2023far, koksal2023longform, wang2022self, honovich2022unnatural, wang2022benchmarking, wei2022finetuned, chung2022scaling}. Both large pre-trained LLMs and human-annotated instruction data are key components of these developments. These LLMs have the ability to perform a wide range of generation tasks. Examples include story writing~\cite{yuan2022wordcraft}, email composition, menu system creation~\cite{kargaran2023menucraft}, poetry writing~\cite{chakrabarty2022help}, code generation~\citep{muennighoff2023octopack}, and food recipe creation~\citep{h2020recipegpt}.

\paragraph{\textbf{Instructional Data Generation.}}
One way to gather instructional data is through the usage of costly human-annotated data~\cite{ouyang2022training}. Additionally, academic NLP tasks can be reformulated to the instructional data format, which often leads to a reduction in the number of text generation tasks~\cite{mishra2022cross, chung2022scaling, wang2022benchmarking}. Another approach involves recent efforts to generate instructional data using LLMs~\cite{honovich2022unnatural, wang2022self}. However, this approach often leads to the creation of datasets with low quality and substantial levels of noise~\cite{wang2022self}. An alternative approach is to use LLMs or pre-defined templates for generating only the instruction part of the data, rather than generating both the instruction and output text pairs~\cite{koksal2023longform}. In this method, the quality of the instructional data is better as only the instruction part is generated, and the output text already has good quality. We use this method later in \secref{instruction-gen} to generate the summary field of \dataname.

\section{\dataname}\seclabel{data}
We now describe \dataname, a dataset in the format of pairs of instructions and corresponding outputs. This dataset comprises a total of \numpairsinstruction data rows.

\subsection{Source Selection} 
There is currently no instructional data available for the task of IRT generation. We use the data provided by \citet{nikeghbal2023girt}, referred to as \girtdata\footnote{\url{https://github.com/ISE-Research/girt-data}} to build our instructional data. \girtdata covers over one million \github repositories, with \numrepoIRT of these repositories supporting IRTs. Each repository can have more than one IRT, and each of these IRTs is accompanied by metadata from the respective repository. We use a part of this data, specifically the IRTs written in Markdown. Most of the repositories prefer to use Markdown. \citet{nikeghbal2023girt} suggested this could be because configuring an IRT in the Markdown format is much easier, and another popular hosting platform \gitlab only uses the Markdown format for IRTs. Thus, in this work, we focus solely on the Markdown part of the data. 
In total, we use \numIRTmarkdown Markdown IRTs from this source.

\subsection{Preprocessing}

We apply preprocessing to ensure data quality so that it can be used for instructional data generation.

\paragraph{\textbf{Filter Null.}}
We observe that some of the IRTs in the \girtdata dataset do not have the \texttt{name} and \texttt{about} fields. These fields are necessary for an IRT to be valid according to \github requirements. We filter out IRTs that do not have these fields. We also observe that some of them do not have a body. Since the most important part of an IRT is the body, we remove IRTs without any bodies. 
After filtering out these IRTs, \numIRTmarkdownnan IRTs remain.

\paragraph{\textbf{Anonymization.}}
We use an anonymized version provided in \girtdata, which renames project names, URLs, etc., with proper tokens. Anonymization helps prevent inadvertent memorization of sensitive information by \modelname during training. When using the \modelname, this anonymization ensures that such information is not generated. However, since only the body part IRT is anonymized in \girtdata, we also apply anonymization to the rest of the IRT as well. We anonymize the \texttt{``assignees''} part of the table which mentions the usernames of developers to whom this type of issue report is assigned, using the \texttt{USER\_i} token. The variable \texttt{i} is assigned a number corresponding to users; for a single user, \texttt{i} is 1, and for multiple users, the first one is assigned 1, the second one is assigned 2, and so forth. We also anonymize the rest of the IRT if the project name appears.

\paragraph{\textbf{Normalization.}} We normalize all of the Markdown IRTs to the same format. Each IRT starts with a table; however, not all of them introduce all the table variables. Additionally, some of them have invalid tables. We use \girtdata metadata on each IRT to complete this table. This is important so the \modelname always produces consistent responses for the IRT table metadata.

\paragraph{\textbf{Deduplication.}} We deduplicate IRTs as some of them are copied from one another without any change. For deduplication, we consider both the body of each IRT and the metadata. We perform exact deduplication, and after that, \numIRTmarkdowndedup IRTs remain. This means that 43\% of IRTs are duplicated from each other without any customization. This number can have two meanings: 
1) Most developers are satisfied with merely copying a template without customization. 2) There is an apparent need for a tool to assist developers in generating customized IRTs. Currently, the most effective way to create an IRT is to explore sample IRTs from other popular repositories, select one that closely matches, and make the necessary adjustments. This process is time-consuming, as it requires checking at least some popular repositories to create a customized version.

\paragraph{\textbf{Filter Non-Latin.}}
We observe that some IRTs are written in languages other than English, especially using CJK characters. Our focus in this work is on English; and, some of the IRT table fields should strictly be written in Latin for an IRT to function properly.

Given the challenges and inaccuracies in language identification, particularly when the text domain differs from the model training data \citep{kargaran2023glotlid}, we rely exclusively on script identification. We eliminate IRTs that are not represented in the Latin script. GlotScript~\citep{kargaran2023glotscript} is used to detect the script of each IRT. After removing non-Latin IRTs, \numIRTmarkdownlatin IRTs remain.

\subsection{Instruction Generation}\seclabel{instruction-gen}

To instruction-tune a language model, our data need to be in the format of instructional data pairs. Each pair comprises an instruction and IRT output where the IRT output corresponds to the given instruction.

In our dataset, the instructions are designed mostly based on the metadata of each IRT. This includes the table information of each IRT (\texttt{name}, \texttt{about}, \texttt{title}, \texttt{labels} and \texttt{assignees}) and structure of IRT (\ie headlines). Also, we have a summary field that provides more specific details about each IRT in natural language.

Our instructions always begin with the metadata of each IRT, we use the anonymized version of preprocessed data for this part.
For the summary field, we use Zephyr-7B-Beta language model~\citep{tunstall2023zephyr} to provide a summary of each IRT.\footnote{We use this prompt to generate summaries: ``You are Zephyr, an AI assistant. Be polite and provide only truthful information. Summarize this \github issue template only using the provided text.''} The objective is to enhance \modelname's capacity to handle developer intentions with more customization in natural language. While this field could be enhanced with annotated data, obtaining annotations is costly and not easily accessible. Many studies augment or generate parts or the entirety of their training examples using language models \citep{koksal2023longform, wang2022self, schick2021generating}. To enable the model to learn both situations (i.e., whether with or without a summary), we generate two sets of instructions: one with the summary field and one without.

\paragraph{\textbf{Masked Instruction.}}
The model should learn to follow the user's instructions. However, users might not always provide answers for all of the information fields in the instructions. For instance, they may not have any suitable suggestions for the label of the IRT. A significant advantage of using language models is their flexibility to generate IRTs while imposing multiple constraints or leaving the choice to the model.
We introduce the summary field, so users can add more details and constraints in natural language. However, the current version of instructions does not have an option to leave the choice to the model for some of the fields. For this purpose, we introduce masked instructions. In this setting, the information fields in the instruction are masked using \texttt{<|MASK|>} token, giving the model flexibility in generating masked fields. So, when a user does not have any idea of what a field should be, using the \texttt{<|MASK|>} token can let the model decide. To train/evaluate the model with this behavior, our instructional data should also include masked instructions. To create masked instructions, we randomly mask the information fields. Each time, we only mask two non-empty fields of the instruction part. The decision to choose two as the number of masked fields is based on the results of a pilot study conducted with the previous version of \modelname, which did not involve masked instructions. In the pilot study, we observed that software developers had trouble filling in an average of two fields. We subsequently added the generated masked instructions to our instructional data.

In total, our dataset includes \numpairsinstruction (\numIRTmarkdownlatin $\times$ 2 $\times$ 2)~<instruction, IRT> pairs. We double our data once by considering summaries and double it again by considering the \texttt{<|MASK|>} token. We randomly split the gathered data into training, validation, and test sets with ratios of 80\%, 10\%, and 10\%, respectively. We ensure that all four instructions assigned to each IRT fall into one of the sets.

\section{\modelname}\seclabel{model}

We now describe \modelname, an open-source model capable of automatically generating IRTs. Users provide their instructions to this model, and \modelname generates an IRT based on the given instruction.

\subsection{Architecture}
We select pre-trained Text-to-Text transfer transformers~(T5)~\citep{raffel2020exploring} model as \modelname base. We use \dataname (\secref{data}) to fine-tune T5 using the HuggingFace library.  
The T5 model is based on the transformer architecture~\citep{vaswani2017attention}, which consists of an encoder-decoder structure. The encoder processes the input text and extracts its features, while the decoder generates the output text.

\paragraph\textbf{{Encoder.}} The T5 encoder is a stack of $N$ identical layers, denoted as $E_1, E_2, ..., E_N$. Each layer applies two sub-layers: a multi-head self-attention mechanism and a feed-forward neural network. The self-attention mechanism captures the contextual dependencies within the input sequence. The feed-forward network introduces non-linearity to the model.

\paragraph\textbf{{Decoder.}} The T5 decoder also consists of $N$ layers, denoted as $D_1, D_2, ..., D_N$. In addition to the self-attention and feed-forward layers, the decoder applies a cross-attention mechanism over the encoder's output. This allows the decoder to attend to different parts of the input text during generation.

\subsection{Fine-Tuning}\seclabel{fine-tune}
To fine-tune the T5 model for the automated generation of IRT, we use a sequence-to-sequence learning approach. The HuggingFace \texttt{Seq2SeqTrainer} optimizes the model parameters by minimizing the loss function, which measures the discrepancy between the generated output and the target output. Now we describe the loss function. We define these parameters:

\begin{itemize}
  \item $D$: Training split of \dataname (<instruction, IRT> pairs).
  \item $x_i$: An input sequence from the dataset.
  \item $y_i$: The corresponding output sequence.
  \item $\theta$: The model parameters (weights) of the T5 model.
\end{itemize}
For a single input-output pair $(x_i, y_i)$, the loss $\mathcal{L}_i$ is defined as follows:

\begin{equation}
\mathcal{L}_i(\theta) = -\sum_{t=1}^{T_{\text{out}}} \log P(y_{i,t} | x_i, y_{i,<t}; \theta)
\end{equation}
Where $T_{\text{out}}$ is the length of the target output sequence $y$, and 
$P(y_{i,t} \allowbreak | x_i, y_{i,<t}; \allowbreak \theta)$ denotes the probability assigned by the model to the token $y_{i,t}$ at position $t$ in the output sequence, given the input $x_i$ and the previously generated tokens $y_{i,<t}$.
Typically, this loss function entails measuring sequence similarity, such as the cross-entropy loss for sequence-to-sequence tasks.

The main goal of fine-tuning is to minimize the average loss across the entire dataset:

\begin{equation}
\mathcal{L}_{\text{fine-tune}}(\theta) = \frac{1}{N} \sum_{i=1}^{N} \mathcal{L}_i(\theta)
\end{equation}

Here, $N$ is the number of input-output pairs in the dataset. 
The loss function penalizes the model for generating incorrect tokens by assigning a higher probability to the target tokens. By optimizing this loss function, the T5 model can learn to generate accurate and contextually relevant output sequences.

\subsection{Tokenizer}
We introduce two types of added tokens to the T5 Tokenizer.

\paragraph{\textbf{Special Tokens.}} 
\dataname uses an anonymized version of IRT. In this anonymized version, any personal details like links have been replaced with appropriate tags. We add all these tags~(i.e., \texttt{<|URL|>}, \texttt{<|Email|>}, \texttt{<|Repo\_Name|>}, \texttt{<|Image|>}) into the special tokens, so they will not be split into subcomponents during tokenization.

We also include the special tokens \texttt{<|EMPTY|>} and \texttt{<|MASK|>} if the user indicates whether a field should be empty or need to be filled by the model.

\paragraph{\textbf{Additional Tokens.}}
 In some language model tokenizers including T5 model tokenizer, there is a normalization process that occurs before tokenization.\footnote{\url{https://stackoverflow.com/questions/72214408/why-does-huggingface-t5-tokenizer-ignore-some-of-the-whitespaces}} This process, by default, removes line breaks represented as \texttt{\textbackslash n}. To ensure that these line breaks are preserved, we add \texttt{\textbackslash n} as an additional token. We also observe the T5 tokenizer generating an unknown token for \texttt{<!---} \texttt{-->} (comment token in markdown), so we include these tokens as additional tokens as well. This ensures that the generated IRT is not compromised by the limitations of a tokenizer that was not trained within this context.

\section{Evaluations Setup}\seclabel{evaluations_setup}

We use the whole test split of \dataname (\secref{data}) for automated evaluations (\secref{automated-evaluation}). 
In automated evaluations, we compare \modelname against baselines (\secref{baselines}).
For human evaluations (\secref{human-evaluation}), we only use a subset (\humantestnum samples) of the test split and only compare \modelname against the best baseline based on automated evaluations.

\subsection{Training Setup}\seclabel{training-setup}

We fine-tune the pre-trained T5-base~\citep{raffel2020exploring} model on the training split of \dataname for 30 epochs using a batch size of 8.
We explain the fine-tuning procedure in \secref{fine-tune}. We use the Adam optimizer~\citep{loshchilov2017decoupled}, with a learning rate of $5e-5$. This optimizer is a common choice for fine-tuning transformer-based models to update the model weights and minimize the loss function.

\subsection{Test Set}\seclabel{test-set}

We use the test split of \dataname (\secref{data}) as our test set. We perform experiments for each type of instruction individually. These types include:

\begin{itemize}
    \item \paragraph{\textbf{META.}} This type includes instructions with the \girtdata metadata.
    \item \paragraph{\textbf{META + MASK.}} This type includes instructions with the \girtdata metadata, wherein two fields of information in each instruction are randomly masked (refer to \secref{instruction-gen} for masked instructions).
    \item \paragraph{\textbf{META + SUM.}} This type includes instructions with the \girtdata metadata and the field of summary.
    \item \paragraph{\textbf{META + SUM + MASK.}} This type includes instructions with the \girtdata metadata and the field of summary. Also, two fields of information in each instruction are randomly masked.
\end{itemize}

\subsection{Baselines}\seclabel{baselines}
We evaluate our \modelname against T5~\citep{raffel2020exploring} and \Flan~\citep{chung2022scaling} on three different sizes (0.2B, 0.8B, and 3B). \Flan is an instruct-tuned version of T5 in a mixture of tasks. We choose \Flan as one of the baselines because it has the same architecture as T5, but it outperforms T5 in various tasks.  Even though they have an equal number of model parameters, Flan-T5 has been trained on a collection of tasks framed as instructions. For evaluation, we use two different prompting techniques: 

\paragraph{\textbf{Zero-Shot.}} Zero-shot prompting evaluates the model's ability to generate IRTs without any examples. For models not trained on the task of IRT generation, models only use pre-trained knowledge. In zero-shot scenarios, the prompt includes only the instruction.

\paragraph{\textbf{One-Shot.}} One-shot prompting involves providing the model with a one-task example in the format of <input, output> pairs, enabling in-context learning. In our one-shot setup, the prompt consists of one pair of <instruction, IRT> followed by the test instruction. The decision to use only one pair is based on the fact that the average length of instructions in \dataname is 128 tokens, and for the IRTs in \dataname, it is 216 tokens. As a result, the mean length of one <instruction, IRT> pair totals 344 tokens. Considering that T5 is trained with a default maximum length of 512 tokens, and we get out-of-memory CUDA errors for the experiments with much longer input sequences, adding extra example pairs is not feasible.

We select one pair of <instruction, IRT> from the most prominent category. Bug report template represent the most frequently occurring type of IRT, accounting for 48.5\% of occurrences in \girtdata. The selected pair in this category is the most frequently used pair that we observed before in \girtdata. We only do one-shot prompting for baselines as the \modelname is trained on the task of IRT generation. We also adjust the format of instruction in the selected <instruction, IRT> for each of the four test set types (see \secref{test-set}).

\subsection{Automated Evaluations}\seclabel{automated-evaluation}

In the absence of specific metrics designed to evaluate IRT generation tasks, we use common text generation metrics. We use three metrics: ROUGE~\citep{lin2004rouge}, METEOR~\citep{banerjee2005meteor}, and BLEU~\citep{papineni2002bleu}, to assess the quality of IRT generation in both zero- and one-shot setups. 

\textbf{ROUGE.} We use ROUGE-1 and ROUGE-L metrics. ROUGE-1 helps to understand how many words overlap and ROUGE-L is the longest matching sequence between the model's output and reference. ROUGE-1 provides an F1 score at the word level. The F1 score is a balance between precision and recall. Precision looks at the correct predictions compared to all predictions, recall assesses the correct inferences compared to all actual samples.

ROUGE-L uses the longest common subsequence method, which takes into account the structure of sentences. This metric automatically identifies the longest chain of words that appear in order, making it sensitive to how words are arranged within sentences, without needing them to be in a strict order.

\textbf{BLEU.} The BLEU metric is widely used to evaluate the quality of machine-generated text. It measures the precision of the generated output by counting the number of overlapping n-grams (contiguous sequences of n items, usually words) between the generated text and the reference(s).

\textbf{METEOR.} The METEOR metric is commonly used for evaluating the quality of text generation. \citet{lavie2004significance} found that metrics emphasizing recall, like METEOR, have a strong focus on matching terms in the model's output with the reference. Importantly, their research shows that these recall-based metrics tend to align better with user preferences than metrics that prioritize precision, such as BLEU.

\begin{table*}[t]
\centering
\captionsetup[figure]{justification=justified, singlelinecheck=off} 
\caption{Automated evaluation of \modelname vs baselines on four test sets. Each test type is represented in one of the blocks.  For example, META refers to instances where instructions are only based on \girtdata metadata (see \secref{test-set} for details). \modelname outperforms all of the baseline models (T5 and \Flan with 3 different sizes) in all of the test sets. The best result per each block is shown in bold, and the second best result is underlined.}\tablabel{automated}
\resizebox{0.99\linewidth}{!}{
\begin{tabular}{l|lr|rrrrrrrr|l|rrrrrrrr}
&
&
& \multicolumn{4}{c}{\textbf{Zero-Shot}}
& \multicolumn{4}{c|}{\textbf{One-Shot}}
&
& \multicolumn{4}{c}{\textbf{Zero-Shot}}
& \multicolumn{4}{c}{\textbf{One-Shot}} \\
& \multicolumn{1}{l}{Model}
& \multicolumn{1}{c|}{Size}
& \multicolumn{1}{c}{ROUGE-1}
& \multicolumn{1}{c}{ROUGE-L}
& \multicolumn{1}{c}{BLEU}
& \multicolumn{1}{c|}{METEOR}
& \multicolumn{1}{c}{ROUGE-1}
& \multicolumn{1}{c}{ROUGE-L}
& \multicolumn{1}{c}{BLEU}
& \multicolumn{1}{c|}{METEOR}

&
& \multicolumn{1}{c}{ROUGE-1}
& \multicolumn{1}{c}{ROUGE-L}
& \multicolumn{1}{c}{BLEU}
& \multicolumn{1}{c|}{METEOR}
& \multicolumn{1}{c}{ROUGE-1}
& \multicolumn{1}{c}{ROUGE-L}
& \multicolumn{1}{c}{BLEU}
& \multicolumn{1}{c}{METEOR}\\
\hline
\midrule
\multirow{7}{*}{\rotatebox{90}{META}} & T5-base & 0.2B 
& 0.05 & 0.05 & 0.00 & 0.02 & 16.83 & 14.60 & 2.23 & 9.55
& \multirow{7}{*}{\rotatebox{90}{META + MASK}} 
& 0.07 & 0.06 & 0.00 & 0.03 & 9.80 & 8.91 & 0.88 & 6.03
\\ \rule{0pt}{2.5ex}
& T5-Large & 0.8B 
& 1.39 & 1.35 & 0.00 & 0.66 & 10.56 & 8.41 & 1.51 & 7.03
&
& 0.83 & 0.78 & 0.00 & 0.30 & 10.74 & 8.27 & 1.77 & 7.42
\\ \rule{0pt}{2.5ex}
& T5-3B & 3B 
& 16.97 & 15.44 & 0.44 & 9.28 & 21.29 & 18.02 & 3.60 & 15.03
&
& 9.92 & 8.95 & 0.33 & 5.98 & 18.03 & \und{15.78} & 2.73 & 13.38
\\ \rule{0pt}{2.5ex}
& \Flan-base & 0.2B 
& 10.42 & 10.12 & 1.61 & 4.14 & 14.83 & 13.05 & 7.51 & 14.43
&
& 3.87 & 3.61 & 1.08 & 2.01 & 5.80 & 5.36 & 3.13 & 5.50
\\ \rule{0pt}{2.5ex}
& \Flan-Large & 0.8B 
& 13.30 & 12.60 & 0.02 & 4.81 & \und{27.95} & \und{21.88} & \und{14.11} & \und{25.36}
& 
& 10.35 & 8.83 & 0.51 & 3.77 & \und{18.23} & 15.18 & \und{7.52} & \und{16.61}
\\ \rule{0pt}{2.5ex}
& \Flan-XL & 3B 
& 9.41 & 9.36 & 0.00 & 3.48 & 18.20 & 16.57 & 9.16 & 15.76
& 
& 3.84 & 3.80 & 0.00 & 1.70 & 8.70 & 7.92 & 3.02 & 7.47
\\ \rule{0pt}{2.5ex}
& \modelname-base & 0.2B 
& \bff{72.34} & \bff{70.20} & \bff{59.96} & \bff{69.95} & \xmark & \xmark & \xmark & \xmark
& 
& \bff{63.44} & \bff{59.74} & \bff{53.93} & \bff{62.95} & \xmark & \xmark & \xmark & \xmark
\\ 
\hline
\rule{0pt}{2.5ex}
\multirow{7}{*}{\rotatebox{90}{META + SUM}} & T5-base & 0.2B 
& 0.09 & 0.08 & 0.00 & 0.04 & 9.07 & 7.49 & 0.42 & 5.86
& \multirow{7}{*}{\rotatebox{90}{META + SUM + MASK}} 
& 0.08 & 0.06 & 0.00 & 0.03 & 10.81 & 9.23 & 0.58 & 6.29
\\ \rule{0pt}{2.5ex}
& T5-Large & 0.8B 
& 5.83 & 4.74 & 0.00 & 2.36 & 10.98 & 8.16 & 3.27 & 7.89
& 
& 8.01 & 6.21 & 0.00 & 3.18 & 11.07 & 8.54 & 3.45 & 8.68
\\ \rule{0pt}{2.5ex}
& T5-3B & 3B 
& 22.43 & 18.72 & 1.06 & 11.90 & 27.41 & 20.86 & 5.26 & 18.03
& 
& 15.51 & 12.34 & 0.48 & 8.16 & 25.47 & 19.81 & 4.21 & 16.76
\\ \rule{0pt}{2.5ex}
& \Flan-base & 0.2B 
& 16.93 & 14.66 & 0.23 & 6.24 & 27.60 & 23.43 & 14.19 & 28.63
&
& 15.23 & 12.53 & 1.02 & 6.03 & 26.51 & 22.18 & 13.27 & 28.19
\\ \rule{0pt}{2.5ex}
& \Flan-Large & 0.8B 
& 17.10 & 13.56 & 0.01 & 6.13 & \und{42.06} & \und{33.28} & \und{20.96} & \und{36.90}
& 
& 13.48 & 10.50 & 0.00 & 4.58 & \und{39.47} & \und{32.26} & \und{21.68} & \und{36.19}
\\ \rule{0pt}{2.5ex}
& \Flan-XL & 3B 
& 17.97 & 15.16 & 0.01 & 7.03 & 18.17 & 15.19 & 6.36 & 15.71
& 
& 16.15 & 12.93 & 0.08 & 6.96 & 20.75 & 17.74 & 9.44 & 17.61
\\ \rule{0pt}{2.5ex}
& \modelname-base & 0.2B 
& \bff{77.90} & \bff{75.33} & \bff{62.94} & \bff{74.38} & \xmark & \xmark & \xmark & \xmark
& 
& \bff{73.92} & \bff{70.32} & \bff{57.12} & \bff{69.85} & \xmark & \xmark & \xmark & \xmark
\end{tabular}
}
\end{table*}

\subsection{Human Evaluations}\seclabel{human-evaluation}

The automated evaluations may not always accurately assess aspects such as consistency and instruction following. We conduct human evaluations to capture these aspects. For our human evaluations, we initially
take the four test sets (refer to \secref{test-set}) and transform each set into n-dimensional TF-IDF vectors. Next, we apply the k-means algorithm to cluster each set into 10 clusters. Subsequently, we randomly take one sample from each cluster for each test set. This approach ensures that the evaluation data are not biased by a single dominant category or instruction type.

We invite \participantnum software engineers who are familiar with the IRT concept to participate in this experiment. We select these maintainers based on our survey from a pool of software engineering volunteers. All of the selected software engineers have been working with version control systems and issue-tracking products for more than five years and are currently using these systems on a daily basis. Participants volunteer for the study and do not receive any compensation. 

We use the IRTs generated by both \modelname and the best model in automated evaluations (\Flan-Large one-shot setup, refer to \secref{automated-result} for more details). We ask each participant to rate all \humantestnum samples without consulting with each other. Additionally, we do not reveal which IRT is created by which models, and we shuffle the order of generated IRTs.

The human evaluations of IRTs considers aspects such as ``consistency and coherence'', ``instruction following'', and ``format'' which involves addressing the following questions:

\begin{itemize}
    \item 
        \textbf{Consistency and Coherence:} Rate the degree of consistency, and coherence within the various segments of the IRTs generated by the two models on a scale from 0 to 5, where 5 represents the highest score.
        For instance, if an IRT starts by focusing on a bug-related topic and then transitions to a non-bug topic, it indicates a poor logical flow.
    \item 
        \textbf{Instruction Following:} Is the generated IRT compliant with the given instruction and does it meet the specified constraints?
        Rate it on a scale from 0 to 5, where 5 represents the highest score.
    \item 
        \textbf{Format:} Does each IRT  conform to the markdown format and \github's IRT standards?~(Yes, Partially, or No)
    
\end{itemize}

\section{Evaluations Results}\seclabel{evaluations_results}

In this part, we discuss the results of experiments outlined in \secref{evaluations_setup}.

\subsection{Automated Evaluations Results}\seclabel{automated-result}

\tabref{automated} provides the results for automated evaluations detailed in \secref{automated-evaluation}. \modelname outperforms all of the baseline models (T5 and \Flan with three different sizes) in all of the test sets. We analyze the results from different aspects in the following.

\paragraph{\textbf{Test Set Type.}} We observe that when instructions contain more details (META + SUM vs META), both \modelname and baselines achieve higher scores.  Additionally, non-masked instructions yield better scores than those with masked tokens (Meta vs META + MASK and META + SUM vs META + SUM + MASK). This difference arises because in masked instructions, the model has greater flexibility in generating the output, but it may be penalized for output mismatches. Nevertheless, the relatively good performance of masked instructions compared to non-masked instructions suggests that even without all the information, the model can still generate IRT to some extent. For instance, when the input instruction for the \texttt{labels} field is masked, the model can still identify the relevant labels based on the context provided in other fields.

\paragraph{\textbf{Instruction-Tuning and Model Size.}} Scaling up language models has been shown to improve performance on a wide range of downstream tasks \citep{wei2022emergent}. This is also the case here. We observe, in most of the cases when increasing the size in T5 and \Flan models, the performance increases. However, there are some cases where the performance worsens. For example, in many setups, \Flan-XL performs worse than \Flan-Large, and the second-best performance in the table is for \Flan-Large. This phenomenon is also observed in some benchmark tasks (e.g., Web of Lies, Tracking Shuffled Objects) in the \Flan paper~\citep{wei2021finetuned}. 

Also, in most cases, the \Flan base models perform better than T5 models of the same size. This is because \Flan is fine-tuned on a collection of tasks, which brings generalization to unseen tasks, such as IRT generation.

\paragraph{\textbf{Zero-Shot vs One-Shot.}} We perform both zero- and one-shot prompting for the baseline models. In all of the setups, one-shot prompting performs better than zero-shot. This is because one-shot prompting enables in-context learning, and the model becomes better aligned with the instruction objective. But still, the best performance among all the general instruction-tuned models with one-shot prompting is worse than \modelname. In addition based on our analysis of a large collection of real-world LLM conversations~\citep{zheng2023lmsys}, we observe, in practice, users interacting with models mostly in a zero-shot setup. So, it's important to have models with the capability of zero-shot setups to help developers who are less familiar with the concept of IRT in the task of IRT generation.

\paragraph{\textbf{Training Progression.}} To analyze the training process, we evaluate \modelname on all instructions available in the validation set. \figref{training_progression} shows that performance improves more rapidly at the beginning of training, but then the rate of improvement slows down. After 25 epochs, the validation loss stabilizes, and the changes in metric values become slightly different. We take a checkpoint of \modelname at the 30th epoch and release it. All experiments are performed using this checkpoint

\begin{figure}[htp]
    \begin{minipage}[t]{0.47\textwidth}
        \centering
        \includegraphics[width=\linewidth]{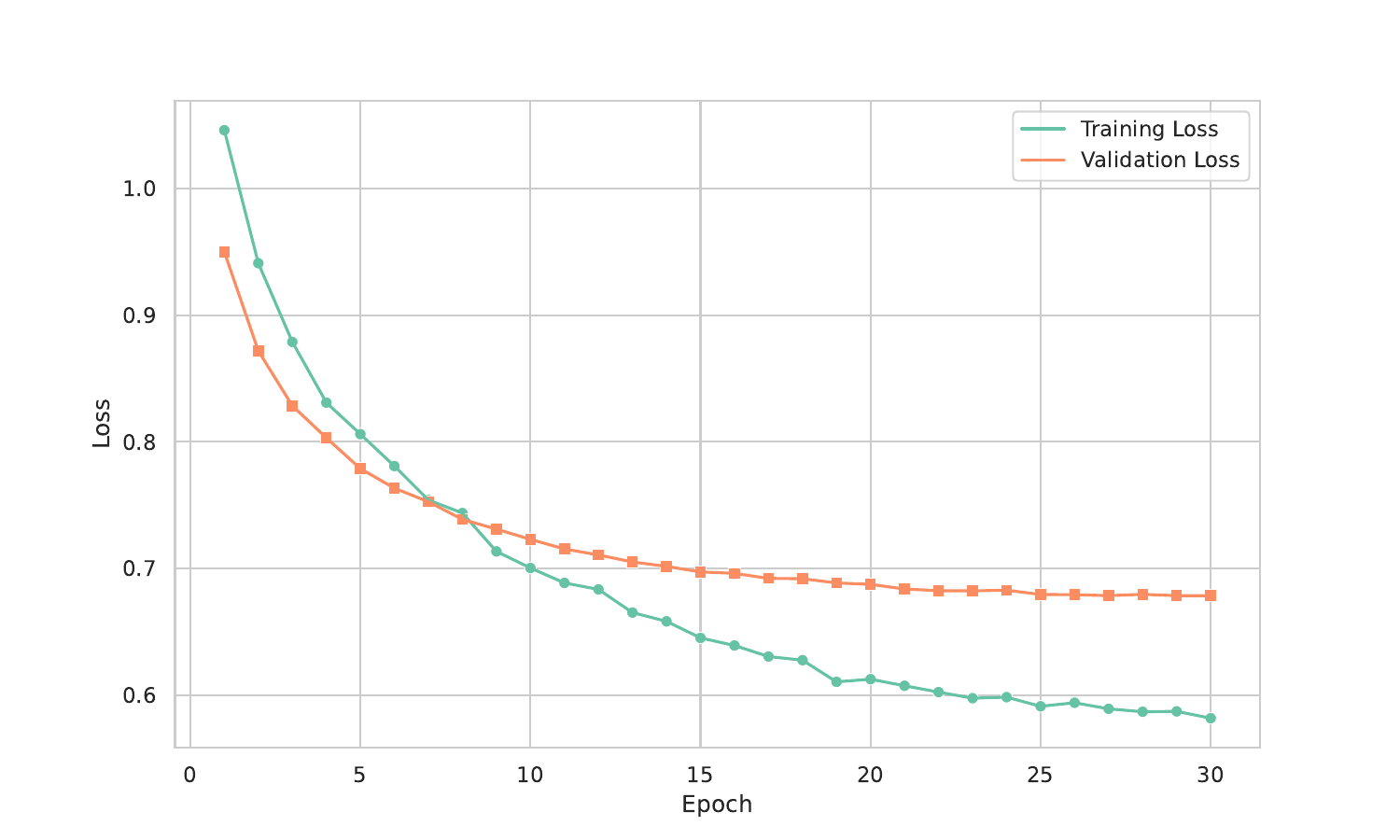}
    \end{minipage}\hfill
    \begin{minipage}[t]{0.47\textwidth}
        \centering
        \includegraphics[width=\linewidth]{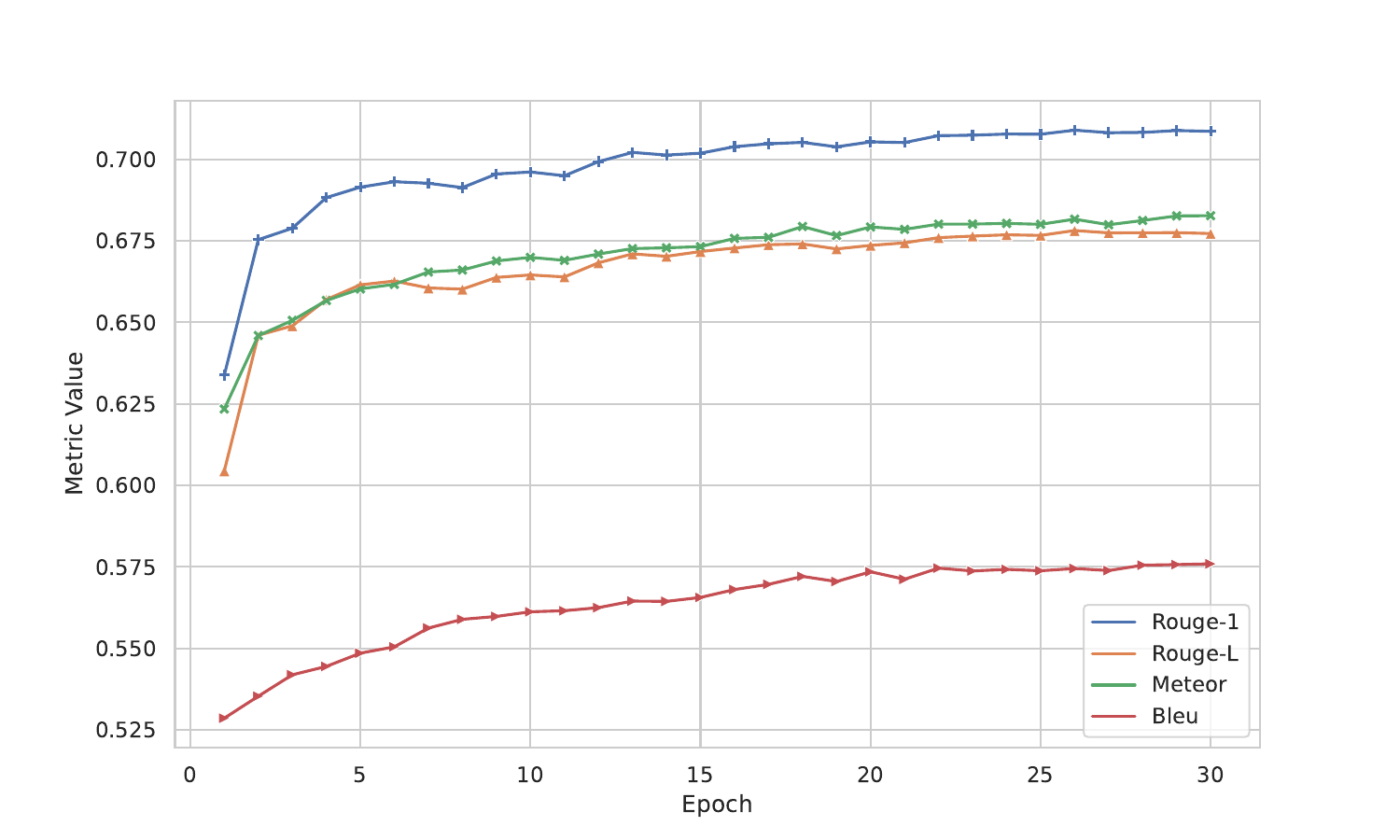}
    \end{minipage}
    \caption[Progression of training]{Top: Progression of training for loss on the validation and training sets. X-axis: Epoch number, Y-axis: Loss value. Bottom: Training progression for metric values on the validation set. X-axis: Epoch number, Y-axis: Metric value.}
    \figlabel{training_progression}
\end{figure}

\subsection{Human Evaluations Results}

\renewcommand{\arraystretch}{1.5} 
\renewcommand{\multirowsetup}{\centering}
\begin{table}[thp]
\centering  
\captionsetup[figure]{justification=justified, singlelinecheck=off} 
\caption{Human evaluation of \modelname (zero-shot) vs \Flan-Large (one-shot). 
\modelname outperforms \Flan-Large in all of the test sets. The scores are normalized to [0, 100\%]. The best result for each test set is shown in bold. }\tablabel{human}
\resizebox{0.99\linewidth}{!}{
\fontsize{8}{7}\selectfont
  \begin{threeparttable} 
    \begin{tabular}{ccrr} 
    \toprule  
    \multirow{2}{1in}{Test Set Type}&  \multirow{2}{0.6in}{Aspect}&\multicolumn{2}{c}{Score \%}\cr
    \cmidrule(lr){3-4}
    &&\Flan-Large & \modelname \cr
    && (One-Shot) \,\,\,\, & (Zero-Shot)  \cr
    \midrule  
    \multirow{3}{1in}{META}
    & Consistency and Coherence & 6.4 & \bff{90.8}\\
    & Instruction Following & 13.6 & \bff{94.0} \\
    & Format (Yes, Partially) &0.0, 4.0& \bff{100.0, 0.0}\cr
    \cmidrule(lr){1-4}
    \multirow{3}{1in}{META + MASK} 
    & Consistency and Coherence & 8.4 & \bff{84.4} \\
    & Instruction Following & 11.2 & \bff{88.4} \\
    & Format (Yes, Partially)&2.0, 6.0 &\bff{100.0, 0.0}\cr
    \cmidrule(lr){1-4}
    \multirow{3}{1in}{META + SUM} 
    & Consistency and Coherence & 24.0 & \bff{91.6} \\
    & Instruction Following & 26.4 & \bff{90.0} \\
    & Format (Yes, Partially)&40.0, 12.0&\bff{100.0, 0.0}\cr
    \cmidrule(lr){1-4}
    \multirow{3}{1in}{META + SUM + MASK} 
    & Consistency and Coherence & 22.4 & \bff{85.2}\\
    & Instruction Following & 24.4 & \bff{88.8} \\
    & Format (Yes, Partially)& 38.00, 2.00& \textbf{100.0, 0.0}\cr
    \bottomrule  
    \end{tabular}
    \end{threeparttable} 
}
\end{table} 

\tabref{human} provides the results of the human evaluations detailed in \secref{human-evaluation}. We normalize 5-scale ratings to [0, 100\%]. Since the best model in automated evaluations results across all test sets is \Flan-Large, for human evaluations, we only compare \modelname with this baseline. \modelname outperforms \Flan-Large in all of the test sets in human experimental evaluation.

Similar to automated evaluations results, we observe that when instructions contain more details (META + SUM vs META), the baseline achieves higher scores. However, this is not completely the case for \modelname, as the ``instruction following'' aspect of evaluation drops when adding summaries. The reason for this could be that with more complex instructions, human evaluators need to assess how well the model adheres to the constraints, and the model may not always fully follow the summary part.

Again similar to automated evaluations, non-masked instructions yield better scores than those with masked tokens (Meta vs META + MASK and META + SUM vs META + SUM + MASK) for both \modelname and \Flan-Large. This is because when the model has more flexibility, it may make mistakes in introducing inconsistencies to the result. On the other hand, the ``instruction following'' aspect is measured more for the non-masked fields since they impose constraints on the output.

Regarding the format, we observe that \modelname completely adheres to the correct format of IRTs. However, since \Flan-Large only sees the format in a one-shot example, it may not consistently follow it correctly. Interestingly, when adding a summary, it follows the example format more than when the summary is not available.

\section{Empirical User Study}\seclabel{sec-user-study}
We conduct an empirical user study to evaluate the effectiveness of \modelname in practice. In this study, \participantnum software engineers, who previously participated in the human evaluations experiment (see \secref{human-evaluation}), are asked to generate IRTs with the \modelname user interface (UI).

\subsection{Empirical User Study Setup}

Each participant is asked to generate one IRT for each of the three distinct issue report types: bug, feature, and question for the Atom project\footnote{\url{https://github.com/atom/atom}}. We select Atom because it is a widely recognized and popular project. Also, its GitHub repository lacks any pre-existing IRTs, ensuring our participants are not influenced by existing templates of this project. Each participant is allocated ten minutes to draft each IRT, which typically consists of 100-200 words. We believe that this duration is sufficient for users to familiarize themselves with the interface and produce suitable IRTs. Participants are informed that they are part of a study about using AI assistance to generate IRTs. However, they are not directly instructed to request assistance from the AI tool. This is done to observe how often users prefer to use AI help. They are also permitted to choose not to use the output of the AI model and are free to use any other resources during the study.

Following the generation of all IRTs, users participate in an exit interview. In this interview, they express their level of agreement in response to the following questions, using a 5-point scale (Strongly Disagree, Disagree, Neutral, Agree, Strongly Agree). Questions include:
\begin{itemize}
    \item \textbf{Usefulness:} How much do you find the AI assistance useful?
    \item \textbf{Ease of Use:} How easy do you find it to use AI assistance in generating IRTs?
    \item \textbf{Goal Achievement:} To what extent are you able to achieve your goals while generating IRTs using AI assistance?
    \item \textbf{Inspiration:} How much does the AI assistant inspire creative ideas for your project's IRTs?
\end{itemize}

Following the scoring of each question, we inquire about the participant's reasons for their scores and feedback. This allows us to improve our model in the future and ensures that participants understand the question correctly.

\begin{figure}[thp]
    \centering
    \includegraphics[width = 0.47\textwidth]{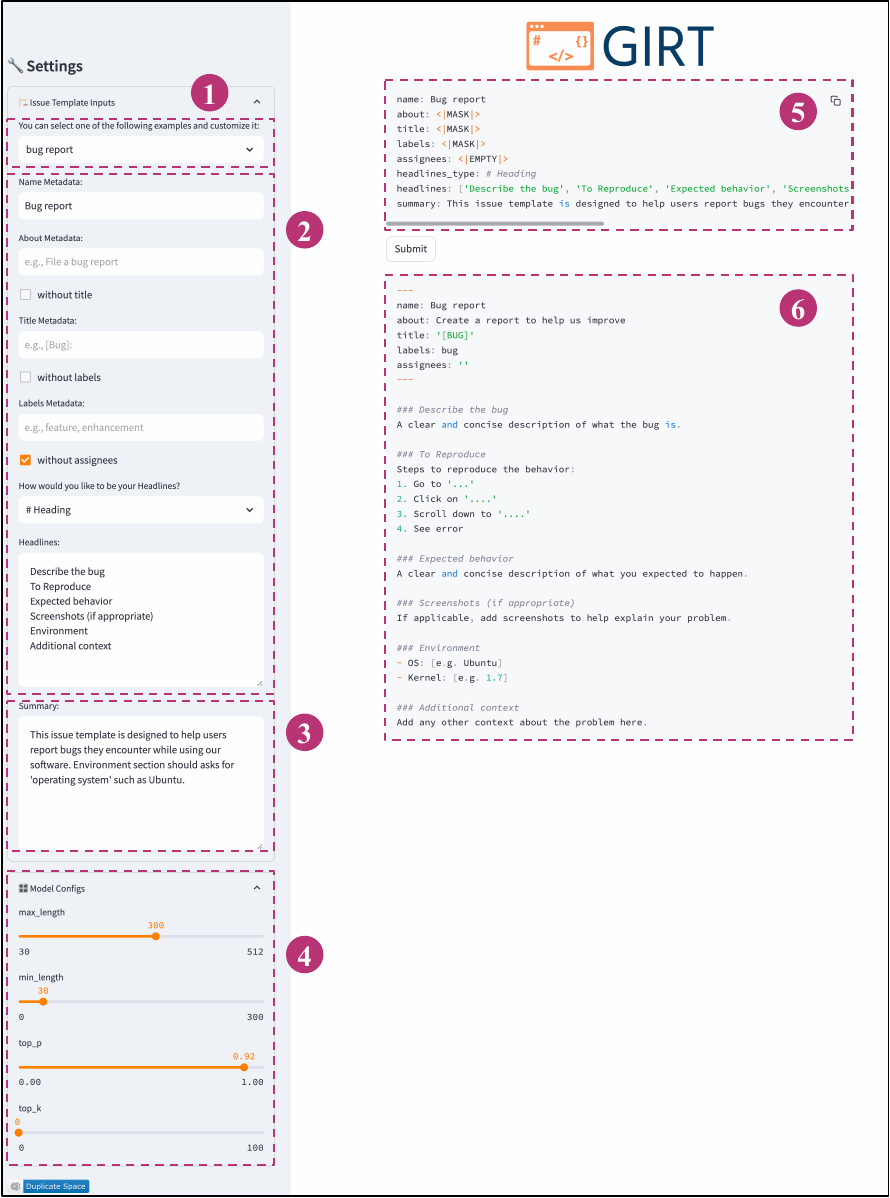}
    \caption[\modelname UI]{UI designed to interact with \modelname. \circledBody{1}{IRT input examples} \circledBody{2}{metadata fields of IRT inputs} \circledBody{4}{summary field of IRT inputs} \circledBody{4}{model config} 
    \circledBody{5}{generated instruction based on the  IRT inputs} \circledBody{6}{generated IRT}. }
    \figlabel{user-interface}
\end{figure}

\subsection{User Interface}

\figref{user-interface} illustrates the \modelname UI. Our aim is to create a user-friendly interface for our experiments, making it simple for users to interact with \modelname. We use the Streamlit library\footnote{\url{https://streamlit.io/}} to design the entire UI and host it on Huggingface space. The model can run on a single CPU core, but for a better user experience, we run it on an NVIDIA T4 during the empirical user study experiment so that users receive the generated IRTs in just a few seconds.

In the UI sidebar, users can input details such as name, about, title, and headlines for the IRT. The summary section allows users to describe their template with more constraints and specify their needs. Users can also choose the empty option for fields they do not want in the generated IRT, \eg title, labels, assignees, or headlines. If they leave a field empty without specifying it should be empty, it will automatically fill with the mask token.

Additionally, there is a ``Model config'' section in the sidebar where users can adjust model settings according to their preferences. These configurations include:

\begin{itemize}
    \item \textbf{max length:}  The maximum number of tokens allowed in the output generated by \modelname. It controls the maximum length of the generated text.
    \item \textbf{min length:} The minimum number of tokens required in the output generated by \modelname. It ensures a certain length for the generated text.
    \item \textbf{top p:} A probability threshold is used in text generation models to limit the sampling pool to the most likely tokens. It helps in focusing on the most probable next words, enhancing coherence.
    \item \textbf{top k:} A fixed number threshold is used in text generation models to restrict the sampling pool to the top-k most likely tokens. Similar to top p, it helps in generating more focused and meaningful text.
\end{itemize}

During an initial test, we observed that some users found it challenging to fill in the inputs. To address this problem, we decided to add some examples to the UI. We want to show the users the types of templates they can generate. However, there is a concern that users might only adjust the current examples. Our primary focus in this work is on the design and evaluation of \modelname. We do not conduct an extensive comparison of different interface designs that could impact human-AI collaboration.  We evaluate the \modelname UI using AIM~\citep{aim2018}, a service and codebase for computational graphical UI evaluation. \modelname receives a ``good'' score for feature congestion~\citep{rosenholtz2007feature, visual_clutter}, indicating that our UI is not cluttered, and it's easy to add new items to the UI. However, regarding the UMSI metric~\citep{fosco2020predicting}, which measures human attention on the UI webpage, our sidebar do not receive enough attention. We plan to improve the UI design based on these measures and user feedback in the future.

\begin{figure}[thp]
    \centering
    \includegraphics[width = 0.47\textwidth]{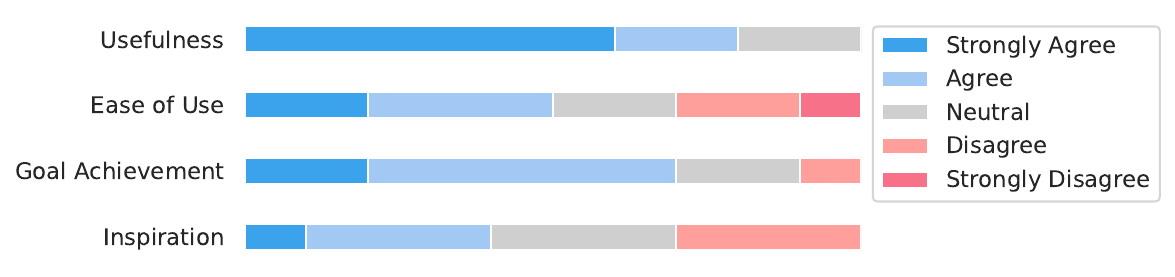}
    \caption{Exit interview results. Participants find \modelname beneficial, reporting high scores for ``usefulness'' and ``goal achievement''.}
    \figlabel{fig-user-study}
\end{figure}

\subsection{Empirical User Study Results}

The results of the empirical user study are shown in \figref{fig-user-study}. Participants highly rate ``usefulness'' and ``goal achievement''. Following that, ``ease of use'' and ``inspirations'' receive the best scores.

The high ratings for ``usefulness'' and ``goal achievement'' indicate that \modelname effectively assists in designing IRTs. According to user feedback, participants report getting bored when trying to write IRTs on their own. Generally, they spend less time generating longer IRTs and achieving their goals with the help of AI assistance.

We expect the ``ease of use'' to receive a lower score compared to ``usefulness'' and ``goal achievement'' since the focus of our work is not involved in creating a best practice UI. For future studies, we plan to invest time in making the interaction within the UI easier.

Regarding ``inspiration'', we do not expect such high scores. Based on participant feedback, several mentioned that simply viewing suggestions from AI assistance is beneficial. It inspires them to generate new ideas to incorporate into their project's IRT, even when they can not use the suggested IRT directly.

\section{Threats to validity}\seclabel{threats}

In this section, we discuss the potential threats to the validity of our model, findings and how
we address or mitigate them. 

\subsection{Internal Validity}
Internal validity relates to the model architecture, hyperparameter settings, and our implementation.

We provide detailed descriptions of our methodology, model architecture, and evaluation process. Additionally, we make our research artifacts, including the model, code, and data, openly available for reproducibility.

\modelname is based on the T5 architecture and employs the default hyperparameter configurations. For training, we report our hyperparameters (refer to \secref{training-setup}). We do not extensively investigate how the T5 parameters impact the performance of \modelname. This decision is made due to the high cost associated with training multiple models. We acknowledge that conducting such a search could enhance the robustness of our findings.

\subsection{External Validity}
Threat to external validity relates to how well our method can be generalized.

\paragraph{\textbf{Effect of Data.}}
In this study, we use the largest publicly available dataset of IRTs, \girtdata, gathered from over one million \github repositories \citep{nikeghbal2023girt}. We use this dataset to have a more diverse representation of IRTs in practical use. We build \dataname, our instructional data, comprising a total of \numpairsinstruction pairs of <instruction, IRT> based on \girtdata. 

The results of our experiments and generalization of \modelname could be limited because of how we construct \dataname. We make an effort to craft instructions in a way that provides users with control over \modelname while maintaining flexibility as needed. We carefully create instructions using a combination of user-entered metadata of desired IRT and summary field. Integrating metadata from \girtdata into instructions acts as a user-friendly guide, offering users the option to personalize each metadata field. To provide the model more flexibility, users can use a \texttt{<|MASK|>} token. This grants the model the freedom to fill in the information for the masked fields based on other contextual details. On the other hand, to limit the model's flexibility and allow users to impose additional constraints, we introduce a summary field as part of the instruction. Users can express their project's IRT requirements in natural language. We use Zephyr-7B-beta LLM to extract the summary from each IRT. We manually inspect some of the generated summaries to make sure they have the expected qualities. However, it's important to note that we cannot guarantee that our instructions are optimal for every IRT. As a result, our findings might be influenced by the instructions used in our experiments. In future research, it would be valuable to investigate how different instructions can impact the outcomes.

\paragraph{\textbf{Effect of Interaction.}}
Interaction with language models through instructions is both powerful and deceptively challenging. Despite their apparent fluency, language models have their own set of conventions and syntax. Crafting effective prompts can be both complex and time-consuming, with sensitivity to word choice, formatting, and exemplar content. Also, language models are highly sensitive to prompt phrasing, and even small changes can lead to significant differences in output. To address these challenges we carefully process all user-entered data (metadata fields and summary field) in our UI to create proper instruction in a stable format that ensures the \modelname performs at its best. 

\paragraph{\textbf{Effect of Pre-trained Language Model.}}
As for the model, we chose T5 as the base for \modelname, which has demonstrated generalization abilities after instruction-tuning for many unseen tasks~\citep{chung2022scaling}.

\subsection{Construct Validity}\seclabel{construct-validity}
Threat to construct validity relates to \allowbreak whether the experimental evaluation and evaluation metrics used in this study are appropriate. In this study, we attempt to include a variety of ways to evaluate \modelname. Our experiments include both automated and human evaluations. Additionally, we conduct a user study to assess \modelname in practice.

Evaluating AI assistance tools can be tricky. It depends on the number of tasks that the tool is capable of, as well as the modality and structure of input-output, and whether gold test data is available. For example, among AI assistance tools, evaluating a tool for the task of grammatical error correction is relatively easy. It's a single task with text as the modality of input and output, without a specific structure, and the gold test data is easy to achieve.

From this point of view, the task of IRT generation is not easy to evaluate. IRTs have a specific structure for which no specific metric is designed to evaluate them. Additionally, the gold data is not available. However, because of IRT text modality, we can use metrics developed for the task of text generation and also employ part of the data (\secref{data}) as the test data. We use popular text generation metrics to assess the performance of \modelname. Specifically, we employ the ROUGE, BLEU, and METEOR (\secref{automated-evaluation}) metrics to evaluate both \modelname and baselines. These metrics are standard for text-generation tasks. However, they cannot capture all aspects of the IRT generation task. For example, do different parts of the generated IRT have coherence, or does the generated IRT adhere to the correct format? For these reasons, we rely on human evaluations to assess \modelname in aspects that automated evaluations do not cover. Additionally, we conduct a user study to check how well the \modelname + UI performs in real-world situations. This helps us understand if the \modelname + UI is practical, user-friendly, and effective in helping users achieve their goals. Moreover, given that the validity of human evaluations and the empirical studies can be significantly influenced by humans, we invited volunteer software engineers familiar with the concept of IRT and who have more than five years of experience with issue-tracking products. 

\section{Limitations}\seclabel{limitation}
Some of the limitations of the proposed approach include:
\begin{itemize}
\item The data for \modelname comes from existing templates available by the developers in \github repositories. \modelname learns and generalizes from these templates; however, its performance is bound by their quality. The motivation behind \modelname is not to surpass the quality of the training data (existing templates). Instead, the proposed model aims to facilitate developers in adopting these templates without the need to search among various \github repositories. It offers increased time savings and customization with less supervision, aligning with the primary objective of this assistant tool.

\item We construct \dataname only by using the metadata of IRTs and a summary generated for the IRTs by the Zephyr-7B-Beta language model. We do not use any other metadata in instructing the instruction part. This work can be enhanced by using additional metadata, such as the topic of the repository, the programming language of the repository, information about the code, dependencies, libraries, and the current issues reported (including their text and categories) to create a more aligned set of IRTs.

\item We use the Zephyr-7B-Beta language model to generate summary fields for each IRT. The data generated by the language model may not always align with how real developers describe the IRT. Acquiring human-annotated data is a costly process. This approach impacts the ``goal achievement'' score in our empirical study; however, based on our observations, users encounter few difficulties in achieving the desired IRT.

\item We use T5 as the base model for \modelname, and we do not explore other architectures and closed-source language models. We are in search of an open-source and cost-effective model specifically designed for this task, aiming for a better developer experience. The entire pipeline of \modelname, including the design of the \dataname (which includes building instructions and masked instruction), evaluations, and the empirical user study, showcases the novelty and possibility of solving this problem using text generation language models.

\item We use the automated evaluations metrics that are not specifically designed for this task. To address the limitations of these metrics, we proposed human evaluations and a user study. Further details on this matter are discussed in construct validity (\secref{construct-validity}).
\end{itemize}

\section{Conclusions and Future Work}\seclabel{conclusion}

Issue reports written using templates are resolved more quickly and align better with developers' expectations. Yet, these templates are not commonly used in repositories, and there is currently no available tool to assist developers in creating them.

To address this problem, we introduced \modelname, an assistant language model that automatically generates issue report templates (IRTs) in the Markdown format. \modelname is specifically designed to generate customized IRTs based on developers' instructions.
We assessed \modelname using both automated and human evaluations. We compared it against six baselines, including T5 and \Flan models with three different sizes. Our experimental results indicated that \modelname performs significantly better than all the baseline models. We also conducted an empirical user study in which we asked software engineer participants to generate IRTs using \modelname's user interface. Participants found \modelname useful, easy to use, inspiring, and highly effective in achieving the intended goals. We hope that through the use of \modelname, we can encourage more developers to adopt IRTs in their repositories.

In the future, we aim to enhance our user interface to make it easier for users to work with \modelname, support the YAML format of IRTs alongside the Markdown format, and also incorporate other metadata such as information about the repository, codebase and currently reported issues into the instruction part of IRTs.

\bibliographystyle{ACM-Reference-Format}
\bibliography{main}

\end{document}